\documentclass[10pt,twocolumn,floatfix,superscriptaddress,amsmath,aps,
pre]{revtex4-1}

\usepackage{xcolor}

\begin{document}

\title{Stochastic dynamics of planar magnetic moments in a three-dimensional environment}
\author{Zochil Gonz\'alez Arenas}
\affiliation{Departamento de Matem\'atica Aplicada, IME, Universidade do Estado do Rio de Janeiro, Rua S\~ao Francisco Xavier 524, 20550-013,  Rio de Janeiro, RJ, Brazil}
\author{Daniel G.\ Barci}
\affiliation{Departamento de F{\'\i}sica Te\'orica,
Universidade do Estado do Rio de Janeiro, Rua S\~ao Francisco Xavier 524, 20550-013,  Rio de Janeiro, RJ, Brazil.}
\author{Miguel Vera Moreno}
\affiliation{Departamento de F{\'\i}sica Te\'orica,
Universidade do Estado do Rio de Janeiro, Rua S\~ao Francisco Xavier 524, 20550-013,  Rio de Janeiro, RJ, Brazil.}
\date{November 25, 2011}

\begin{abstract}
We  study the stochastic  dynamics of a two-dimensional magnetic moment  embedded in a three-dimensional environment,  described by means of the 
stochastic Landau-Lifshitz-Gilbert (sLLG) equation. We define a covariant 
generalization of this equation, valid  in the ``generalized Stratonovich discretization prescription''. We present  a path integral formulation that allows to compute any 
$n-$point correlation function, independently of the stochastic calculus used. Using this formalism, we show the equivalence between the 
cartesian formulation with vectorial noise, and the polar formulation with just one scalar fluctuation term.  In particular, we show  that, 
for isotropic fluctuations, the system is represented by an {\em additive stochastic process}, despite of the multiplicative terms appearing in the original formulation of the sLLG equation, but, for anisotropic fluctuations the noise turns out to be truly multiplicative. 
\end{abstract}

\maketitle

%%%%%%%%%%%%%%%%%%%%%%%%%%%%%%%%%%%%%%%%%%%%%%%%%%%%%%
\section{Introduction}
%%%%%%%%%%%%%%%%%%%%%%%%%%%%%%%%%%%%%%%%%%%%%%%%%%%%%%

The technology for manufacturing two-dimensional magnetic materials has been increasingly developed in the last two 
decades~\cite{OHandleyBook2000}. The importance of these materials runs from the understanding of their basic physical 
behaviour to the development of novel devices. 
Magnetic nanoparticles, confined in two dimensional geometries, display a variety of magnetic textures that have been 
observed in several compounds using a wide range of  experimental 
techniques~\cite{MagneticNano-Fu1985, MagneticNano-Berger1992, MagneticNano-Pescia1998, MagneticNano-Puntes2004}. In particular, 
magnetic stripe structures have been observed in FeCu films~\cite{VaStMaPiPoPe2000}  that were interpreted as nematic 
phases originated by the competition of short distance ferromagnetic interactions and long distance dipolar 
interactions~\cite{BaSt2007,BaSt2009,StBa2010,BaSt2011,BaRiSt2013,MeStNi2015,MCBaSt-2017}.
Generally, in-plane long distance ferromagnetic correlations are elusive since, due to the Mermin-Wagner theorem,  
thermal fluctuations destroy ferromagnetic order.
However, two-dimensional ferromagnetism can be effectively considered in anisotropic compounds, 
where thermal fluctuations are keeping in check~\cite{2DFerro-Gong2017}.

These facts motivate us to study dynamics of two-dimensional magnetic moments in a ferromagnetic environment, embedded in an 
anisotropic three-dimensional sample.  Below the Curie transition temperature,  it is possible to define a local order parameter 
$\vec m(x,t)$, consisting in the expectation value of the magnetization per unit volume. Usually, this macroscopic quantity is 
described by the so called micro-magnetic approach, modelled by the 
stochastic Landau-Lifshitz-Gilbert equation (sLLG)~\cite{Brown1963,Kubo1970}, Eq.~(\ref{eq.sLLG}) below.   
This equation describes the dynamics of a classical magnetization  with constant modulus. The only relevant variable here is
the orientation of the local magnetization. The dynamics is driven by external forces, by damping effects and by randomly distributed 
magnetic field fluctuations. 

The sLLG is a highly nontrivial stochastic nonlinear equation and has been applied to many particular situations with very  
different methods, analytical as well as numerical 
ones~\cite{GarciaPalacios1998, MagneticNano-Hanggi2007,Billoni2007,Cimrak2008,Bertotti2009,Aron2014,Roma2014,Nicolis2016-1,Nicolis2016-2}.  
From a mathematical point of view, the sLLG equation is a system of stochastic equations with multiplicative noise~\cite{vanKampen,Gardiner}. 
For multiplicative white noise, it is necessary to fix a discretization scheme to correctly interpret its meaning. 
The most popular schemes are It\^o and Stratonovich prescriptions, however there are in principle infinite possible prescriptions 
that define not only the specific 
form of the stochastic equation but also the calculus rules necessary to deal with them. 
In particular, the discretization scheme of the sLLG has raised some controversy, due to the fact that the stochastic evolution must 
keep the modulus of the magnetization constant~\cite{GarciaPalacios1998,Berkov2002}.
We have recently introduced a functional path integral formalism~\cite{arenas2010,arenas2012} to deal with multiplicative processes, that 
does not depend on the particular calculus used to define the Langevin equation. 
This formalism is very useful, in particular, to establish fluctuation and work relations~\cite{ArBaCuZoGus2016} 
that involve time reversal transformations. Indeed, we have shown that  time reversal transformations connect different discretization 
prescriptions~\cite{Arenas2012-2,Miguel2015} and, for this reason, it is interesting to have a technique which deals with any 
prescription in a unified formalism.

In this work we study the sLLG equation by using a class of discretization schemes called 
``generalized Stratonovich prescription"~\cite{Hanggi1978} or ``$\alpha$-prescription"~\cite{Janssen-RG}. Each discretization 
scheme is parametrized by a real number $0\le\alpha\le 1$, in such a way that the usual It\^o and Stratonovich prescriptions are particular 
cases ($\alpha=0$ and $\alpha=1/2$, respectively). Moreover, the prescription $\alpha=1$ is known as the 
H\"anggi-Klimontovich interpretation~\cite{Hanggi1978, Hanggi1980, Hanggi1982, Klimontovich} or 
anti-It\^o prescription, which is the time reversal conjugate to the It\^o interpretation~\cite{Miguel2015}. 
We consider three different approaches for dealing with the sLLG equation, using its particular advantages and relating them. These approaches
are Langevin and Fokker-Planck equations and the path integral formalism.

Specifically, we present a ``covariant''  version of the sLLG equation which conserves the constant modulus of the magnetization for any 
discretization prescription. To do this, we interpret the prescription changes as a gauge transformation and define an 
``$\alpha$--covariant time derivative" to produce prescription independent observables. As a consequence, an interesting  geometrical character of 
the discretization schemes is explicitly displayed. In this paper, we focus on a two-dimensional projection of the covariant sLLG, 
keeping the three-dimensional character of the noise. We show that the dynamics of this system is described by a simple covariant 
multiplicative Langevin equation for an angular variable with just one noise variable. For the case of uncorrelated magnetic fluctuations, 
the stochastic process reduces to a {\em Markov additive process}. Conversely, for correlated noise, the process is intrinsically 
multiplicative  producing an equilibrium distribution that deviates from the usual Boltzmann type. 

The paper is organized as follows: In section~\ref{sec.model}, we review some important mathematical properties of the sLLG equation.  
Then, we present the covariant version of this equation in the $\alpha$-prescription and, finally, we project the magnetization dynamics to 
two dimensions. In \S~\ref{sec.FP} we analyse equilibrium properties by solving the associated Fokker-Planck equation in some particular cases. 
In section~\ref{sec.PI} we built up a path integral formalism for the two-dimensional model and show that most of the properties described 
in previous sections are actually valid for any $n$-point correlation function.
Finally, we discuss our results and future perspectives in section~\ref{sec.Discussion}.

%%%%%%%%%%%%%%%%%%%%%%%%%%%%%%%%%%%%%%%%%%%%%%%%%%%%%%%%%%%%%%%%%%%%
\section{Stochastic dynamics of planar magnetic moments}
\label{sec.model}
%%%%%%%%%%%%%%%%%%%%%%%%%%%%%%%%%%%%%%%%%%%%%%%%%%%%%%%%%%%%%%%%%%%%
We begin this section by reviewing some mathematical properties of the three-dimensional stochastic sLLG equation. 
The dynamics of a classical local magnetization $\vec m(t)$, with constant modulus $|\vec m(t)|=m_0$, in contact with a medium 
modelled by a friction term and a noisy thermal bath, is described by  the following equation:
\begin{equation}
\frac{d\vec{m} }{dt}= \gamma \vec{m}\times \left[\vec {\cal B }+\vec b \right]-\frac{\gamma\lambda}{m_0}
\vec{m}\times\left(\vec{m}\times \left[\vec {\cal B} +\vec b \right]\right) \; .
\label{eq.sLLG}
\end{equation}
The first term of Eq.~(\ref{eq.sLLG}) is the classical drift produced by a local magnetic field ${\cal B}$, while the second 
term models the dissipation in the medium. $\gamma$ is the gyromagnetic ratio and $\lambda$ is a dimensionless
damping coefficient. The local magnetic field  $\vec {\cal B }$ is given by an effective Hamiltonian
\begin{equation}
\vec {\cal B}(\vec m) =-\frac{\partial H_{eff}}{\partial \vec m} \; .
\end{equation}
Moreover, the thermal bath is modelled by magnetic field fluctuations, $\vec b$,  randomly distributed according to
\begin{eqnarray}
\langle b_i(t)\rangle&=&0   
\label{eq.meanb0}\\
\langle b_i(t)b_j(t')\rangle&=&2 D_{ij} \delta(t-t')   
\label{eq.cn}
\end{eqnarray}
where $i,j=1,2,3$ and the correlation matrix $D_{ij}$ is, in principle, an arbitrary $3\times 3$ constant matrix. Later, we will 
specify physically reasonable matrix structures for the problem we focus on in this paper. 
Eq.~(\ref{eq.sLLG}) was originally formulated~\cite{Brown1963} to describe  a  three dimensional local magnetization in an isotropic environment, 
in which the correlation $D_{ij}=\delta_{ij}$. In this work, we consider a more general noise distribution to take into account anisotropic 
materials. 
To completely define equation~(\ref{eq.sLLG}), 
it is necessary to specify a prescription to discretize the equation, since the products $\vec m\times \vec b$ or $\vec m\times(\vec m\times \vec b)$ 
are ill-defined due to the fact that $\vec b(t)$ is delta correlated. 
Eq.~(\ref{eq.sLLG}) has to be interpreted  in the Stratonovich convention for which, under time discretization with 
intervals  $(t_j,t_{j+1})$ with $j=1,2\ldots$, we evaluate the magnetization $\vec m$ at  an intermediate time $\tau_j$ given, 
at each interval, by
\begin{equation}
\vec m(\tau_j)=\frac{1}{2}\left[\vec m(t_j)+\vec m(t_{j+1})\right]  \; .
\label{eq.StrPrescription}
\end{equation} 
This prescription is very useful to make analytical mani\-pu\-lations since the calculus rules are the usual ones. However, sometimes, it could be convenient to work in other stochastic prescriptions. 
For instance, in numerical computations, it is easier to consider a ``causal" pre-point, or the It\^o prescription in which 
 $\vec m(\tau_j)=\vec m(t_j)$ is chosen.  In general, any stochastic differential equation can be discretized in the so-called 
 ``generalized Stratonovich prescription''~\cite{Hanggi1978} 
or ``$\alpha$-prescription''~\cite{Janssen-RG}, for which 
\begin{equation}
\vec m(\tau_j)=(1-\alpha)\vec m(t_j)+\alpha \vec m(t_{j+1})\mbox{~~ with~~} 0\le \alpha \le 1. 
\label{eq.prescription}
\end{equation}
In this way, $\alpha=0$ corresponds with the pre-point It\^o interpretation and $\alpha=1/2$ coincides with the (midpoint) Stratonovich one.  
The post-point prescription, $\alpha=1$,  is the time reversal conjugate of the It\^o prescription~\cite{arenas2012,Miguel2015}. 
Each particular choice of $\alpha$ fixes a different stochastic evolution. Interestingly, it also fixes  different calculus rules for each prescription.  

In order to have a deeper insight,  it is convenient to re-write Eq.~(\ref{eq.sLLG}) in a more standard form, 
generally considered for any multiplicative stochastic system,  
\begin{equation}
\frac{d m_i}{dt} =f_i(\vec m)+g_{ij}(\vec m) \eta_j
\label{eq.standar}
\end{equation}
with delta-correlated noises
\begin{eqnarray}
\langle\eta_i(t)\rangle&=& 0 
\label{eq.meaneta0}\\
\langle\eta_i(t)\eta_j(t')\rangle&=&\delta_{ij}\delta(t-t') \; .
\label{eq.wn}
\end{eqnarray}
$f_i(\vec m)$ is a vector drift force and the diffusion matrix  $g_{ij}(\vec m)$  defines the multiplicative character of the  noise.

By simple algebraic manipulations, we can  show that Eq.~(\ref{eq.standar}), with the noise distribution Eqs.~(\ref{eq.meaneta0})  
and~(\ref{eq.wn}),
are exactly equivalent to Eq.~(\ref{eq.sLLG}), with the noise distribution Eqs.~(\ref{eq.meanb0})  and~(\ref{eq.cn}),  provided we identify, 
\begin{eqnarray}
f_i&=& \gamma m_0\Gamma_{ij} {\cal B}_j 
\label{eq.fm} \\
g_{ij}&=& \gamma m_0 \Gamma_{i\ell} A_{\ell j} \sqrt{2D_{(j)}}. \;
\label{eq.gm}
\end{eqnarray}
The matrix $\Gamma$ in Eqs.~(\ref{eq.fm}) and~(\ref{eq.gm}) can be split into a symmetrical ($\Gamma^{(s)}$) and an 
anti-symmetrical ($\Gamma^{(a)}$) part 
$\Gamma=\Gamma^{(a)}+\lambda\Gamma^{(s)}$, whose components are given by
\begin{eqnarray}
\Gamma_{ij}^a&=&-\epsilon_{ijk} \hat m_k 
\label{eq.Gammaa} \\
\Gamma_{ij}^s&=&\delta_{ij} -\hat m_i\hat m_j \; ,
\label{eq.Gammas}
\end{eqnarray}
where $\epsilon_{ijk}$ is the completely antisymmetric Levi-Civita tensor and  $\hat m_i$ are the components of the unit 
vector $\hat m=\vec m/m_0$.
In Eq.~(\ref{eq.gm}), the matrix $A$ diagonalizes the correlation matrix $D_{ij}$, 
\begin{equation}
A^T_{i\ell} D_{\ell m} A_{mj}=D_{(i)} \delta_{ij}
\end{equation}
where  $D_{(i)}$ are the eigenvalues of $D$.
Interestingly, the matrices $\Gamma^{(a)}$ and $\Gamma^{(s)}$ are transversal projector operators, $m_i\Gamma^{(a)(s)}_{i,j}=0$, 
having the following  algebraic properties,   
\begin{equation}
\Gamma^{(a)T}\Gamma^{(a)}=\Gamma^{(s)},\;\Gamma^{(s)T}\Gamma^{(s)}=\Gamma^{(s)},\;\Gamma^{(s)T}\Gamma^{(a)}=0 .
\label{eq.algebra}
\end{equation}

Consider, for instance, a stochastic vector variable $\vec m(t)$ satisfying Langevin Eq.~(\ref{eq.standar}), interpreted
in the $\alpha$-prescription. As already mentioned, the discretization scheme not only modifies the temporal evolution, but also modifies 
the calculus rules when dealing with the stochastic variable $\vec m(t)$. 
In particular, in order to compute total time derivatives of any function $F(\vec m(t))$, the stochastic chain rule reads~\cite{Miguel2015} 
 \begin{eqnarray}
 \lefteqn{
\frac{d F(\vec m(t))}{dt}= \frac{\partial F}{\partial m_j} \frac{d m_j}{dt}+
\frac{(1-2\alpha)}{2}\frac{\partial^2 F}{\partial m_i\partial m_j}
g_{ik} g_{jk} }  \; \;
\label{eq.ChainRule} \\
&=&\frac{\partial F}{\partial m_j} \frac{d m_j}{dt}+
(1-2\alpha)\gamma^2 m_0^2\frac{\partial^2 F}{\partial m_i\partial m_j}
\Gamma_{ik} D_{k\ell}\Gamma_{j\ell} \;,
 \nonumber
\end{eqnarray}
where in the second line we have used the explicit expression for $g_{ij}$ given by Eq.~(\ref{eq.gm}).
 Choosing $F=|\vec m|^2$, we  can immediately compute the total time derivative of the magnetization modulus
\begin{equation}
\frac{d |\vec m(t)|^2}{dt}= 2  m_j \frac{d m_j}{dt}+
2(1-2\alpha) \gamma^2 m_0^2\Gamma_{ik} D_{k\ell} \Gamma_{i\ell} \; .
\end{equation}
Using transversality, $m_i \Gamma_{ij}=0$, as well as  the algebraic properties Eq.~(\ref{eq.algebra}) we find, 
\begin{equation}
\frac{d |\vec m(t)|^2}{dt}=
2(1-2\alpha)(1+\lambda^2)\gamma^2 m_0^2\;  {\rm Tr}(\Gamma^{(s)} D), 
\label{eq.magnetmodulus}
\end{equation}
where we have used matrix notation.

In Eq.~(\ref{eq.magnetmodulus}), we see that only for $\alpha=1/2$, $d|\vec m|^2/dt=0$. Therefore, only the Stratonovich prescription  
keeps the modulus of the magnetization constant. Interestingly, while in deterministic theory, transversality is the only 
requirement to keep the modulus constant, in stochastic evolution, it is necessary  to choose the correct discretization 
prescription in addition  to transversality. For this reason, the original work on sLLG~\cite{Brown1963} is formulated in the 
Stratonovich prescription scheme.  
This si\-tua\-tion produced some misunderstanding in the literature~\cite{GarciaPalacios1998,Kamppeter1999,Berkov2002}, 
specially due to numerical computations with difficulties in keeping $|\vec m|=m_0$ constant. 

There are several reasons to work with other prescriptions different from Stratonovich. For instance, the ``natural" 
discretization for the implementation of computer algorithms is the pre-point It\^o prescription ($\alpha=0$), due to its 
causal character. Moreover, the anti-It\^o prescription ($\alpha=1$) is interesting when dealing  with fluctuation theorems, 
in which it is necessary to implement time reversal stochastic evolutions. Thus, whether  we decide to work with any prescription 
$\alpha\neq 1/2$, Eq.~(\ref{eq.sLLG}) should be modified  in such a way that the modulus of the magnetization remains constant. 
The canonical way to do that is to translate the drift $f_i(\vec m)$. 
To be precise, consider a stochastic  mean value of any function of  the stochastic variable ${\vec m(t)}$, computed by using 
Eq.~(\ref{eq.standar}), interpreted in the Stratonovich sense. Then, the {\em same mean values} can be computed by means of the equation  
\begin{equation}
 \frac{dm_i}{dt} = \left[f_i+\frac{(1-2\alpha)}{2}g_{kj}\partial_k g_{ij} \right] + 
g_{ij}({\bf x})\eta_j \ , 
\label{eq.translated}
\end{equation}
interpreted in the $\alpha$-prescription. In Eq.~(\ref{eq.translated}) and in the rest of the paper 
$\partial_k\equiv \partial/\partial m_k$. 
We can represent the \emph{same stochastic process} by means of {\em different stochastic differential equations}, by just 
shifting the drift through
\begin{equation}
f_i({\vec  m})\to f_i({\vec m})+\frac{(1-2\alpha)}{2} g_{k\ell}({\vec m})\partial_k g_{i\ell}({\vec  m}).
\label{eq:Translator}
\end{equation}
Each differential equation is  interpreted with a  different discretization prescription. Moreover, the calculus rules are also 
changed for each stochastic differential equation.

A detailed explanation of Eq.~(\ref{eq:Translator}) can be found in Ref.~\cite{Miguel2015} (specially in the appendix for 
rigorous demonstrations).  
Sometimes, this translation was called ``spurious drift term"~\cite{Lubensky2007}, and it was used to force the convergence to the Boltzmann 
equilibrium distribution for any value of $\alpha$. The present context is somewhat different, since the drift was introduced to  
guarantee that all stochastic differential equations with different values of $\alpha$ represent the same stochastic process, where 
the modulus of the magnetization is constant. 
Interestingly, there is a special condition on the diffusion matrix in which the drift translation is not necessary (\emph{i.e.}, it is zero). 
If, for instance~\cite{Hanggi1978}, 
\begin{equation}
g_{k\ell}({\vec m})\partial_k g_{i\ell}({\vec m})=0,
\label{eq:gdg}
\end{equation}
then the evolution is independent of the prescription. We will show that, while  this is not true for three-dimensional magnetic moments, 
it is  the case for planar magnetic rotors under uncorrelated noise ($D_{ij}=\delta_{ij}$). 

Due to the algebraic properties of the $\Gamma$ matrices  
there is a much more elegant and deeper way to understand the ``spurious drift term''.  By means of purely algebraic manipulations, it is 
not difficult to show that 
\begin{equation}
g_{k\ell}({\vec m})\partial_k g_{i\ell}({\vec m})=-\frac{1}{m_0^2} {\rm Tr}\left(g^Tg\right)\, m_i \,.
\end{equation}
In this way, Eq.~(\ref{eq.translated}) can be rewritten in the following form, 
\begin{equation}
D^{(\alpha)}_t m_i =f_i(\vec m)+g_{ij}(\vec m) \eta_j \;,
\label{eq.GsLLG}
\end{equation} 
where  the operator $D_t^{(\alpha)}$ is the {\em $\alpha-$covariant derivative}, given by
\begin{equation}
D^{(\alpha)}_t=\frac{d~}{dt}+\frac{(1-2\alpha)}{2m_0} Tr(g^T g) \; .
\label{eq.CovariantDerivative-G}
\end{equation} 
Or, in terms of the $\Gamma$ matrices, 
\begin{equation}
D^{(\alpha)}_t=\frac{d~}{dt}+(1-2\alpha)(1+\lambda^2)\gamma^2 Tr(\Gamma^{(s)} D) \; .
\label{eq.CovariantDerivative}
\end{equation} 

This form of the equation allows a deeper geometric interpretation of the connection between different prescriptions. Indeed, 
the second term of Eq.~(\ref{eq.CovariantDerivative-G}) or~(\ref{eq.CovariantDerivative}) behaves as a gauge field and the drift 
translation between different $\alpha$ is, in fact, a gauge transformation. The covariant derivative is enforcing the fact that, 
under changes in the discretization scheme, the stochastic differential equation changes, however, without modifying the underlying 
stochastic process, \emph{i.e.}, any correlation function is $\alpha$--independent. 
A caveat is appropriate here. We are using the terminology of {\em gauge transformations} in a statistical sense, \emph{i.e.}, under changes 
in the discretization prescription, the detailed stochastic trajectories indeed change, however, the stochastic averages 
(the observables) are not modified. 

Notice that $D^{(\alpha)}_t$ has the correct properties of a time derivative. On one hand, the gauge term has inverse time units, 
since  $[\gamma^2 D]=[t]^{-1}$. On the  other, it is odd under time reversal. The reason for this is that 
a stochastic time reversal transformation involves a change in the discretization prescriptions. Particularly, the time reversal operator is given 
by~\cite{Miguel2015}
\begin{equation}
{\cal T} = \left\{
\begin{array}{lcl}
{\vec m}(t) &\to & {\vec m}(-t)           \\   & & \\
\alpha &\to & (1-\alpha) \hspace{3cm},      \\ & & \\
f_i &\to&  f_i  +\left(2\alpha-1\right)\; g_{k\ell}\partial_k g_{i\ell}
\end{array}
\right.
\label{eq.T}
\end{equation}
in such a way that ${\cal T}^2={\cal I}$ and  ${\cal T} D^{(\alpha)}_t=-D^{(\alpha)}_t$, as it should be.

Due to the transversality of $\Gamma$, it is immediate to show that Eq.~(\ref{eq.GsLLG}) implies the covariant relation,   
\begin{equation}
m_i D^{(\alpha)}_t m_i =0 \;.
\end{equation}
From this property and  using the generalized chain rule Eq.~(\ref{eq.ChainRule}),  we show that 
$d |\vec m|^2/dt=0$ for any value of $\alpha$. As a consequence, Eq.~(\ref{eq.GsLLG}) is the correct covariant generalization of 
the sLLG equation for arbitrary stochastic prescriptions and for correlated magnetic field fluctuations.

In this paper, we are interested in highly anisotropic systems, in such a way that the magnetic moment is confined to live in a plane. 
In these systems, the only relevant dynamical variable is an angle, since we can write 
$\vec m=m_0(\cos\theta, \sin\theta)$ in the $x-y$ plane. 
We can implement this projection by considering a  highly anisotropic Hamiltonian that penalizes any  out of plane 
magnetization or, equivalently, by imposing $m_z=0$ as a hard constraint. 
However, since the planar magnet is embedded in a three-dimensional environment, we will keep the three components of the noise, $b_i$. 
A reasonable, although general, correlation matrix $D_{ij}$, is
\begin{equation}
D_{ij}=\left(
\begin{array}{ccc}
 D+\Delta &  \delta & 0 \\
 \delta & D-\Delta & 0 \\
 0 & 0 &  D_{\perp}
 \end{array}
\right).
\label{eq.D}
\end{equation}
The structure of  $D_{ij}$ takes into account the strong anisotropy in the $z$-axis, and  we have also allowed a weak noise anisotropy in the 
$x-y$ plane, given by the value  of $\Delta<< D$, and  weak planar correlations,  measured by $\delta<<D$.  

In order to rewrite Eq.~(\ref{eq.GsLLG}) in terms of $\theta$, we need to make a nonlinear change of variables. This is a bit 
cumbersome to perform for any value of the prescription $\alpha$, since we need to use the generalized chain rule, Eq.~(\ref{eq.ChainRule}). 

Let us firstly show how to make a projection in the simpler case of the Stratonovich prescription, and afterwards we present 
the covariant equation valid in any discretization scheme.  Replacing  $\vec m(t)=m_0(\cos\theta(t), \sin\theta(t))$ into Eq.~(\ref{eq.GsLLG}), 
fixing $\alpha=1/2$ and using usual calculus rules, we find a simpler stochastic differential equation given by 
\begin{equation}
\frac{d\theta}{dt}=  f(\theta) + \vec{g}(\theta)\cdot \vec b(t)
\label{eq.3noise}
\end{equation}
where
\begin{eqnarray}
f(\theta) &=& \gamma\lambda \left[{\cal B}_y \cos\theta-{\cal B}_x \sin\theta\right] \nonumber \\
&=&-\frac{\gamma\lambda} {m_0} \frac{\partial H_{eff}}{\partial \theta} \, ,
\label{eq.ftheta}
\end{eqnarray}
and  
\begin{eqnarray}
g_x(\theta)&=&-\gamma\lambda \sin\theta ,
\label{eq.gx} \\
g_y(\theta)&=&\;\;\; \gamma\lambda \cos\theta ,
\label{eq.gy} \\
g_z(\theta)&=&-\gamma.
\label{eq.gz}
\end{eqnarray}
The noise satisfies the white noise correlated distribution $\langle b_i(t) \rangle=0$, $\langle b_i(t)b_j(t') \rangle=2D_{ij}\delta(t-t')$, 
where $D_{ij}$ is given by Eq.~(\ref{eq.D}).
Therefore, the system is reduced to a Langevin equation with three noise terms, one additive ($g_z$) and two 
multiplicative ($g_x(\theta),g_y(\theta)$). 

This system is completely equivalent to the 
usual multiplicative noise Langevin equation, with just one delta-correlated noise term, 
\begin{equation}
\frac{d\theta}{dt}=  f(\theta) +  g(\theta)\eta\, ,
\label{eq.1noise}
\end{equation}
with $\langle \eta(t)\eta(t')\rangle=\delta(t-t')$, $f(\theta)$ unchanged  (given by Eq.~(\ref{eq.ftheta}))
and $g(\theta)$ such that
\begin{equation}
 g^2(\theta)= \sqrt{2} \gamma^2\left(D_\perp+\lambda^2\left\{D-\Delta\cos(2\theta)-\delta\sin(2\theta)\right\} \right).
\label{eq.g}
\end{equation}
In section~\ref{sec.PI} we will show, using path integrals,  that the equivalence between Eq.~(\ref{eq.3noise}) and the system given by 
Eq.~(\ref{eq.1noise}) is valid for any n-point correlation function.

While Eq.~(\ref{eq.1noise}) was deduced in the Stratonovich prescription, it is also possible to deduce the covariant equation valid  
for any value of $\alpha$. Starting from 
Eq.~(\ref{eq.GsLLG}) and performing the nonlinear change of variables, using the proper generalized chain rule to make the transformation 
and then reducing the three noises to one term we find, 
\begin{equation}
D_t^{(\alpha)}\theta=  f(\theta) +  g(\theta)\eta \, ,
\label{eq.G1noise}
\end{equation}
where the representation of the covariant derivative acting on the phase $\theta(t)$ is, 
\begin{eqnarray}
D_t^{(\alpha)} \theta&=& \frac{d\theta}{dt}+\frac{1}{2}(2\alpha-1) g g'  
\label{eq.Covarianttheta} \\
&=&  \frac{d\theta}{dt}+\frac{1}{2}(2\alpha-1) \gamma^2\lambda^2 \left(\Delta \sin(2\theta)+\delta \cos(2\theta)\right).
\nonumber
\end{eqnarray}
As it is usual in gauge theories, the specific form of the covariant derivative depends on the representation of the field to which 
it applies. Eq.~(\ref{eq.Covarianttheta}) has the general form  $D_t\theta=d\theta/dt+A_0$, resembling  the structure of a gauge covariant 
derivative acting on a periodic phase field like, for instance, in $XY$ models or even superconducting models.   

Summarising, Eq.~(\ref{eq.G1noise}) with the covariant derivative given by Eq.~(\ref{eq.Covarianttheta}) is the appropriate stochastic equation 
describing planar magnetization in a three-dimensional environment, valid for any discretization scheme parametrised by $0<\alpha<1$.

In the usual treatments of the sLLG equation, the noise is considered uncorrelated. In this case, $\Delta=\delta=0$ in Eq.~(\ref{eq.D}). 
Therefore, from Eq.~(\ref{eq.g})  we see that the noise 
is  $\theta$ independent and so, the stochastic process is an \emph{additive Markov process} with the diffusion constant given by 
\begin{equation}
 g = \sqrt{2} \gamma \sqrt{D_\perp+\lambda^2 D} \; .
 \label{eq.gconstant}
\end{equation}
However, for correlated noise it is not true and the stochastic process is truly multiplicative.

%%%%%%%%%%%%%%%%%%%%
\section{Equilibrium properties}
\label{sec.FP}
%%%%%%%%%%%%%%%%%%%%
Following the methods of Ref.~\cite{arenas2012}, we compute,  from Eq.~(\ref{eq.G1noise}),  a Fokker-Planck equation for the 
probability distribution $P(\theta,t)$:
\begin{equation}
\frac{\partial P}{\partial t}=-\frac{\partial}{\partial \theta }\left\{f+\frac{1}{2} gg'\right\}P+
\frac{1}{2}\frac{\partial^2}{\partial \theta ^2}\left(g^2 P\right),
\label{eq.FP}
\end{equation}
where $g'=dg/d\theta $.
This equation is $\alpha$-independent, confirming that the covariant Langevin equation 
Eq.~(\ref{eq.G1noise}) correctly describes $\alpha$ independent observables (stochastic mean values).

Eq.~(\ref{eq.FP})  can be cast into a  continuity equation, 
\begin{equation}
\frac{\partial P(\theta ,t)}{\partial t}+\frac{\partial J(\theta ,t)}{\partial \theta }=0,
\label{eq.Continuity}
\end{equation} 
with the probability current given by
\begin{equation}
J(\theta ,t)= \left[f-\frac{1}{2} gg'\right] P(\theta ,t)- \frac{1}{2}g^2 \frac{\partial P(\theta ,t)}{\partial \theta }.
\label{eq.J(x,t)}
\end{equation} 
The differential equation should be supplemented with the periodic condition $P(\theta,t)=P(\theta+2\pi, t)$ and the 
normalization $\int_0^{2\pi} d\theta P(\theta,t)=1$. 

We suppose  that, at long times, the probability  converges to a steady
state $P^S(\theta)$,  given by 
\begin{equation}
P^S(\theta)=\lim_{t\to \infty}P(\theta,t)=N\; e^{-U(\theta)},
\end{equation} 
with the normalization constant $N^{-1}=\int_{0}^{2\pi} d\theta\; e^{-U(\theta)}$. In this state, the stationary current is
\begin{equation}
J^S(\theta)= N e^{-U(\theta)}\;\left(f-\frac{1}{2} gg'+\frac{1}{2}g^2 \frac{dU(\theta)}{d\theta}\right)
\label{eq.JS}
\end{equation} 
and the stationary Fokker-Planck equation acquires the simpler form 
\begin{equation}
\frac{dJ^S(\theta)}{d\theta}=0,
\label{eq.stationaryFP}
\end{equation} 
or, simply, $J^S(\theta)=\bar J= \mbox{constant}$. 
Let us focus on the equilibrium state, defined as the solution of the
stationary Fokker-Planck equation with zero current probability. Thus,
for $\bar J=0$, there is an obvious solution of
equations~(\ref{eq.JS}) and (\ref{eq.stationaryFP}), given by 
\begin{equation}
U_{\rm eq}(\theta)=-2 \int^\theta \frac{f(\theta')}{g^2(\theta')} d\theta'+ \frac{1}{2} \ln g^2(\theta).
\label{eq.equilibriumpotential}
\end{equation}
While the first term of  Eq.~(\ref{eq.equilibriumpotential}) is the contribution of the drift force  to the equilibrium potential, 
the second term is a pure noise contribution, coming from the multiplicative character of the stochastic process.
We show some interesting equilibrium properties in the following particular cases.
%%%%%%%%%%%%%%%%%
\subsection{Particular cases}
\subsubsection{Isotropic noise}
In the usual uncorrelated noise case ($\Delta=0,\delta=0$), $g(\theta)\equiv \mbox{constant}$, given by Eq.~(\ref{eq.gconstant}), and  the 
stochastic process is additive. The  last term 
of Eq.~(\ref{eq.equilibriumpotential})  can be absorbed in the normalization factor.   Then, the equilibrium distribution is given by 
\begin{equation}
U_{\rm eq}= \frac{2\lambda}{m_0\gamma  (D_\perp+\lambda^2 D)} H_{eff} \;.
\label{eq.Uadd}
\end{equation}
The fluctuation-dissipation theorem imposes the Einstein relation 
\begin{equation}
\frac{2\lambda}{m_0\gamma (D_\perp+\lambda^2 D)} =\beta=\frac{1}{k_B T} 
\label{eq.fd}\, ,
\end{equation}
with $k_B$ the Boltzmann constant and $T$ the temperature. 
In this way, the long time distribution probability has the usual Boltzmann form  $P_{\rm eq}\sim e^{-\beta H_{eff}}$.

For concreteness, consider a simple Hamiltonian describing an isotropic planar
magnetic moment in the presence of a magnetic field pointing in the x-direction,
\begin{equation}
H_{eff}= -\vec m\cdot \vec B= -m_0 B_x \cos\theta.
\end{equation}
In this case, the effective Langevin equation is equivalent to a stochastic
pendulum~\cite{Gitterman2008},
\begin{equation}
\frac{d\theta}{dt}= -\gamma\lambda B_x \sin\theta +\eta \, ,
\label{eq.sPendulum}
\end{equation}
with $\langle \eta(t)\eta(t')\rangle= g^2\delta(t-t')$, and the long time equilibrium potential takes the form 
\begin{equation}
U_{\rm eq}(\theta)= -\frac{2\lambda B_x}{\gamma  (D_\perp+\lambda^2 D)}\;  \cos\theta \;.
\label{eq.Upendulum}
\end{equation}

\subsubsection{Anisotropic noise}

The situation is quite different for correlated and/or anisotropic noise, where the noise is effectively multiplicative. 
In this case, the equilibrium probability is not a Boltzmann distribution. The reason for this is inherent to multiplicative noise, which has 
a more involved interpretation  than additive noise.  In an additive process, it is clear that the noise is ``external'', 
in the sense that, in the absence of noise, we have a deterministic problem well defined by a Hamiltonian.  Conversely, 
in multiplicative noise, or ``internal noise''  this is not always possible. In fact, the presence of  noise profoundly modifies the 
original ``classical''  system. In other words, the splitting of the model in two parts, one ``classical'' or ``deterministic'' 
and the  other ``fluctuating''  is too naive. 

To illustrate this concept, let us consider a Hamiltonian describing a simple anisotropy in the $x$-axis, 
\begin{equation}
H_{eff}=-h m_x^2 = -h m_0^2 \cos^2\theta\; , 
\label{eq.xanisotropy}
\end{equation}
where $h$ measures the intensity of the anisotropy. We will take  $0<\Delta<D$ and $\delta=0$ to describe noise correlations with 
essentially the same type of anisotropy as the Hamiltonian.
The equilibrium potential can be easily computed from Eq.~(\ref{eq.equilibriumpotential}), obtaining,
\begin{equation}
U_{\rm eq}(\theta)=\left[\frac{ m_0 h+\lambda \gamma\Delta}{2\lambda\gamma \Delta}\right]\ln\left(1-\frac{\lambda^2\Delta}{D_\perp+\lambda D}\cos 2\theta
\right) \; .
\label{eq.Uanisotropic}
\end{equation} 
Evidently, $U_{\rm eq}$ is not proportional to $H_{eff}$, leading to a probability distribution quite different from the Boltzmann type. 
Of course, if we compute the limit of zero $\Delta$, in which the noise turns out to be additive,  we obtain
\begin{equation}
\lim_{\Delta\to 0} U_{\rm eq}=\beta H_{eff} \, ,
\end{equation}
consistently with  Eqs.~(\ref{eq.Uadd}) and~(\ref{eq.fd}).
We can gain more insight from this example by Taylor expanding $U_{\rm eq}$ in powers of 
$\Delta/(D_\perp+\lambda^2 D)$. To linear order we find, 
\begin{eqnarray}
U_{\rm eq}&\sim& \beta H_{eff}
\label{eq.expansion} \\
&-&\frac{1}{2}\left(\frac{\lambda^2 \Delta}{D_\perp+\lambda^2 D}\right)
\left\{\cos 2\theta+\frac{h m_0\lambda}{2\gamma(D_\perp+\lambda^2 D) }\cos 4 \theta\right\}.
\nonumber 
\end{eqnarray}
The first line of Eq.~(\ref{eq.expansion}) is the Boltzmann limit for $\Delta=0$, while the second line is the linear correction 
in $\Delta$. We see that it has the contribution of higher harmonics of the anisotropy that cannot be cast in the original form of the 
Hamiltonian. The greater $\Delta$, the greater the number of harmonics contributing to the equilibrium potential. 

Moreover, lets look a bit closer to Eq.~(\ref{eq.expansion}). Making $h\to 0$ for a fix value of $\Delta$ we get that, while $H_{eff}=0$, 
the equilibrium potential takes the form
\begin{equation}
\label{eq.hzero}
U_{\rm eq}\sim
 -\left(\frac{\lambda^2 \Delta}{D_\perp+\lambda^2 D}\right) \; 
\cos^2\theta\; .
\end{equation}
Using the definition of temperature of Eq.~(\ref{eq.fd}), we get
\begin{equation}
\label{eq.hzero2}
U_{\rm eq}\sim
 -\left(\frac{\lambda  m_0 \Delta}{2}\right) \beta m_0^2 \; 
\cos^2\theta \;.
\end{equation}
This potential  has essentially the same form of Eq.~(\ref{eq.xanisotropy}), in which  $\Delta$ is taking the role of an effective 
anisotropic field, $h_{eff}\sim \lambda m_0 \Delta/2$. Therefore, we have essentially the same equilibrium potential for two very 
different situations: in the former case, we have  an anisotropic Hamiltonian subjected to isotropic additive noise. In the latter case ($h=0$),
we have a   purely noised system, ($H_{eff}=0$),  and a small anisotropic noise ($\Delta<<(D_\perp+\lambda^2D)$).  This simple example shows up  
the difficulty of splitting the effects of the deterministic contribution, given by $H_{eff}$, from the noise contribution, 
given by $g(\theta)$. In general, both contributions are intertwined, producing 
equilibrium potentials far away from the Boltzmann form (such as, for instance, Eq.~(\ref{eq.Uanisotropic})).

%%%%%%%%%%%%%%%%%%%%%
\section{Path integral approach}
\label{sec.PI}
%%%%%%%%%%%%%%%%%%%%%%

In this section, we discuss the path integral formalism, appropriated to represent $n$-point correlation functions 
of two-dimensional 
magnetic rotors embedded in a three-dimensional environment. In addition to the power of the formalism as a calculation 
tool~\cite{Zinn-Justin,TirapeguiBook1982,WioBook2013}, it is very 
useful to study symmetries and  its corresponding Ward-Takahashi identities that give rise to fluctuation theorems 
in equilibrium~\cite{Arenas2012-2} as well as out of equilibrium~\cite{ArBaCuZoGus2016}.  

The main motivation of this section is to show the exact equivalence between the three-noise vector model of Eq.~(\ref{eq.standar}) 
and the scalar equation with just one noise variable which defines the angular model of Eq.~(\ref{eq.1noise}). That means that all 
the $n-$point correlation functions are the same in both formulations. On the other hand, it will become evident that the two-dimensional 
stochastic process is additive when the noise in uncorrelated and isotropic; however, it is nontrivially multiplicative in the presence 
of even a small correlation between components of the noise or any anisotropy.
For simplicity,  we will work in this section in the Stratonovich prescription. The generalization to any other prescription can 
be done following Refs.~\cite{Miguel2015,Aron2014}.

Here, we apply the general methods of Ref.~\cite{Miguel2015}, to the present problem of two-dimensional magnets in a three-dimensional environment. 
We want to build up a representation of  n-point correlation functions, defined as
\begin{equation}
\langle m_{i_1}(t_1)\ldots m_{i_n}(t_n)  \rangle\equiv\langle  m_{bi_1}(t_1)\ldots m_{bi_n}(t_n)  \rangle_{\vec b}\; .
\label{eq.def}
\end{equation} 
Here, $i_1,i_2,\ldots,i_n=1,2$ are the components of the two-dimensional magnetization $m_i(t)$.  $\vec m_b$ are solutions of 
Eq.~(\ref{eq.sLLG}) in two dimensions, for a given realization of the noise $\vec b$. $\langle \ldots\rangle_{\vec b}$ means stochastic 
average with respect to the Gaussian distribution, Eqs.~(\ref{eq.meanb0}) and~(\ref{eq.cn}),  for the vectorial noise $\vec b$. 
The main idea is that these correlation functions can be computed by functional deriving a ge\-ne\-rating functional in the following way
\begin{equation}
\langle m_{i_1}(t_1)\ldots m_{i_n}(t_n)  \rangle=\left.\frac{\delta^n Z[\vec J]}{\delta J_{i_n}\ldots\delta J_{i_1}}\right|_{\vec J=0}\; .
\label{eq.deltaZ}
\end{equation} 
The generating functional  is defined as
\begin{equation}
Z[J] = \langle e^{\int dt \vec J(t)\cdot \vec m_b}\rangle\, ,
\label{eq.Z0}
\end{equation}
where, as before,  $\vec m_b$ is a solution of Eq.~(\ref{eq.sLLG}), for a par\-ti\-cu\-lar
realization of the noise $\vec b(t)$. $\vec J(t)$ is a localized two-dimensional source and  $\langle\ldots\rangle$ represents 
stochastic averages with respect to the three-dimensional noise $\vec b(t)$. It is straightforward  to show that 
Eq.~(\ref{eq.deltaZ}), with the $Z[J]$ functional defined by Eq.~(\ref{eq.Z0}),  is formally equivalent to Eq.~(\ref{eq.def}).

We want to compute the stochastic average in Eq.~(\ref{eq.Z0}), in order to find a representation of $Z[J]$ that does not 
depend on the explicit solution of Eq.~(\ref{eq.sLLG}). To achieve this goal, the first step is to define an auxiliary function 
$\vec m(t)$, by  introducing a functional Dirac delta distribution in the following way, 
\begin{equation}
e^{\int dt \vec J(t)\cdot \vec m_b}=\int {\cal D} \vec m\; \delta^2\left( \vec m-\vec
m_b\right) e^{\int dt \vec J(t)\cdot \vec m}  \ .
\label{eq.expmb}
\end{equation}
With this trick, we move the explicit solution $\vec m_b(t)$ from the exponential to the functional delta. Now, we use the following property,
\begin{equation}
\delta^2\left( \vec m-\vec m_b\right)=\delta^2\left( \hat O(\vec m)\right) \det\left(\frac{\delta \hat O}{\delta \vec m}\right)  ,
\label{eq.deltaproperty}
\end{equation}
where the operator $\hat O(\vec m_b)=0$. This property is the functional generalization of the well-known property of the single 
variable Dirac delta function $\delta(f(x))=\delta(x-x_0)/|f'(x_0)|$, with $f(x_0)=0$ and $f'$ the total derivative with respect to $x$.
The last term of Eq.~(\ref{eq.deltaproperty}), ${\cal J}(\vec m)=\det(\delta \hat O/\delta\vec m)$, is simply the Jacobian of the variable 
transformation 
$\vec m\to \hat O(\vec m)$.

Replacing Eq.~(\ref{eq.deltaproperty}) into Eq.~(\ref{eq.expmb}) and Eq.~(\ref{eq.Z0}) we can write the generating functional as   
\begin{equation}
Z[J] =  \int  {\cal D} \vec m\; \left\langle\delta^2\left( \hat O(\vec m)\right) {\cal J}(\vec m)\right\rangle\; 
e^{\int dt \vec J(t)\cdot \vec m} \;.
\label{eq.Z1}
\end{equation}

The important achievement of this Eq.~(\ref{eq.Z1}) is that $Z[\vec J]$ does not depend anymore on the explicit 
solution $\vec m_b$ of the sLLG equation. So, the stochastic average is only applied to the functional delta and the Jacobian, 
since they are the only noise-dependent objects in the expression.

From Eq.~(\ref{eq.sLLG}), it is natural to define the vector operator $\hat O(\vec m)$ as
\begin{equation}
\hat O(\vec m)\equiv\frac{d\vec{m} }{dt}-\gamma \vec{m}\times \left[\vec {\cal B }+\vec b \right]+\frac{\gamma\lambda}{m_0}
\vec{m}\times\left(\vec{m}\times \left[\vec {\cal B} +\vec b \right]\right) \; .
\label{eq.Om}
\end{equation}
With this choice, we guarantee that $\hat O(\vec m_b)=0$.
Taking advantage of the transversality of Eq.~(\ref{eq.sLLG}), it is more convenient to work with the radial ($\vec m\cdot \hat O$) 
and transversal ($\vec m\times \hat O$) components  of this equation. Thus, we define the scalar function $O_r$ and the 
pseudo scalar $O_\theta$ as
\begin{eqnarray}
O_r(\vec m)&\equiv&  \vec m \cdot \frac{d\vec{m} }{dt} ,\\
O_\theta(\vec m) &\equiv& \vec m \times \frac{d\vec{m} }{dt}-\vec m\times \vec F,
\end{eqnarray}
with 
\begin{equation}
\vec F= \gamma \vec{m}\times \left[\vec {\cal B }+\vec b \right]-\frac{\gamma\lambda}{m_0}
\vec{m}\times\left(\vec{m}\times \left[\vec {\cal B} +\vec b \right]\right)\; .
\end{equation}
The equations $O_r(\vec m_b)=0$ and $ O_\theta(\vec m_b)=0$ are completely equivalent to Eq.~(\ref{eq.sLLG}).

For the sake of simplicity, we write $O_r$ and $O_\theta$ in polar coordinates, $\vec m(t)= m(t)(\cos\theta(t),\sin\theta(t))$, obtaining 
\begin{eqnarray}
O_r&=&  m \frac{dm}{dt}
\label{eq.Orpolar}\, , \\
O_\theta&=&
m^2 \left\{\frac{d\theta}{dt}-\left(f(\theta)+\vec g(\theta)\cdot \vec b(t) \right)\right\} \, ,
\label{eq.Othetapolar}
 \end{eqnarray}
 with $f(\theta)$ and $\vec g(\theta)$ given by Eqs.~(\ref{eq.ftheta}) to~(\ref{eq.gz}). 
 We can now rewrite the generating functional,  Eq.~(\ref{eq.Z1}),  as
\begin{equation}
Z[J] =  \int  {\cal D} \vec m\; \left\langle \delta\left(O_r\right)
\delta\left(O_\theta\right) \; {\cal J}(\vec m) \right\rangle\; \;  e^{\int dt \vec
J(t)\cdot \vec m} \, ,
\end{equation}
where the Jacobian is 
\begin{equation}
{\cal J}(\vec m)=\det\left(
\begin{array}{ccc}
 \frac{\delta  O_r}{\delta m} &  &  \frac{\delta  O_r}{\delta \theta} \\
  & & \\
 \frac{\delta  O_\theta}{\delta m} & &  \frac{\delta O_\theta}{\delta
\theta}
 \end{array}
\right).
\end{equation}
Due to  the fact that $\delta  O_r/\delta \theta=0$ (see Eq.~(\ref{eq.Orpolar})), the determinant factorizes as 
\begin{equation}
{\cal J}=\det\left(\frac{\delta  O_r}{\delta m}\right)\;
\det\left(\frac{\delta  O_\theta}{\delta
\theta}\right) \; . 
\end{equation}
The functional integral can then be written as
\begin{align}
Z[J] &= \int  {\cal D}m  {\cal D}\theta\;  m(t) \;\det\left(\frac{\delta O_r}{\delta m}\right) \delta\left(O_r\right)  \nonumber \\ 
&\times \left\langle \delta\left(O_\theta\right) \; \det\left(\frac{\delta O_\theta}{\delta
\theta}\right)\right\rangle\; \;  e^{\int dt \vec J(t)\cdot \vec m} \; ,
\end{align}
where we have used the functional measure in polar coordinates and we have taken advantage from the fact that  
$O_r$ is noise-independent; for this reason, it is not affected by the stochastic average. The effect of $\delta(O_r)$ is 
simply to enforce the constraint $|\vec m|=m_0$. This is a direct consequence from $\vec F\perp \vec m$.

After integration in ${\cal D}m$ we have, up to normalization constants,  
\begin{align}
Z[J] = & \int {\cal D}\theta \;\;  \left\langle \delta\left(O_\theta\right)  \det\left(\frac{\delta  O_\theta}{\delta
\theta}\right)\right\rangle\nonumber \\
& \times e^{ m_0\int dt\,  \left(J_x(t) \cos\theta+J_y(t) \sin\theta\right)} \; .
 \label{eq.z3}
\end{align}

The next step is to compute the stochastic average. To do this we represent the functional delta by its functional Fourier transform, 
\begin{equation}
\delta\left(O_\theta\right)=\int {\cal D}\varphi \; \; e^{- i\int dt\; \varphi O_\theta},
\label{eq.varphi}
\end{equation}
where we have introduced the so-called response auxiliary variable $\varphi(t)$.

The determinant can be also exponentiated by means of a couple of conjugate Grassmann variables $(\bar\xi,\xi)$~\cite{arenas2010} 
in the following way, 
\begin{equation}
\det\left(\frac{\delta O_\theta}{\delta
\theta}\right)=\int  {\cal D}\bar\xi {\cal D}\xi\;\; e^{\int dt\; \bar\xi 
\left(\frac{\delta  O_\theta}{\delta \theta}\right)  \xi} .
\label{eq.xi}
\end{equation}

Using Eqs.~(\ref{eq.varphi}) and~(\ref{eq.xi}) and the explicit expression for $O_\theta$, Eq.~(\ref{eq.Othetapolar}), we have 
 \begin{align}
 &\left\langle \delta\left(O_\theta\right) \; \det\left(\frac{\delta O_\theta}{\delta
\theta}\right)\right\rangle=\label{eq.smean} \\
& \int {\cal D}\varphi{\cal D}\bar\xi {\cal D}\xi \;e^{-i m_0^2\varphi\left\{\frac{d\theta}{dt}-f\right\} +m_0\left\{\bar\xi\frac{d\xi}{dt}-f'\bar\xi\xi\right\}} 
\left\langle   e^{\int dt\; \sum_i P_i b_i} \right\rangle  ,
\nonumber
\end{align}
where 
\begin{equation}
 P_i=m_0^2\left(i\varphi g_i - g'_i \bar\xi\xi\right) ,
\end{equation}
with $f'=df/d\theta$ and $g'_i =dg_i/d\theta$.

Thus, the effect of the exponentiation of the functional delta and the Jacobian, is to quote the noise content of the generating functional as 
an exponential of a {\em linear function of the noise $\vec b$}.  Considering that the noise distribution 
is Gaussian, it is immediate to compute the stochastic average by using the following expression
\begin{equation}
\left\langle   e^{\int dt\; \sum_i P_i b_i} \right\rangle=e^{\int
dt\; \sum_{ij} P_i D_{ij} P_j} \,,
\label{eq.Gmean}
\end{equation}
where $i=x,y,z$ and $D_{ij}$ are given by Eq.~(\ref{eq.D}).

 Explicitly computing the quadratic form we find 
\begin{equation}
 P^T D P=-m_0^4\left\{\frac{1}{2}\varphi^2(t) g^2(\theta)- i\varphi g(\theta) g'(\theta)\bar\xi(t)\xi(t) \right\},
\end{equation}
where we have
\begin{equation}
 g^2(\theta) = \sqrt{2} \gamma\left[D_\perp+\lambda\left\{D-\Delta\cos(2\theta)-\delta\sin(2\theta)
\right\}\right],
\label{eq.g2}
\end{equation}
which exactly coincides with Eq.~(\ref{eq.g}).

Now, replacing these results into Eq.~(\ref{eq.z3}) and rescaling  variables, 
$\varphi\to \varphi/m_0^2$, $\xi\to \xi/m_0$ and $\bar\xi\to \bar\xi/m_0$,  we find for the generating functional
\begin{equation}
 Z(J)=\int {\cal D} \theta{\cal D} \varphi {\cal D} \bar\xi{\cal D} \xi\;
e^{-S(\theta,\varphi,\bar\xi,\xi)+\int dt (J_x\cos\theta+J_y\sin\theta)} \; ,
\end{equation}
where the action reads
\begin{equation}
 S = \int dt \left\{\frac{1}{2} \varphi^2 g^2 + i \varphi\left(\frac{d\theta}{dt}-f +  g g'\bar\xi \xi \right) - \bar\xi \frac{d\xi}{dt}+f' \bar\xi\xi \right\}.
\end{equation}
Notice that the modulus of the magnetization $m_0$ dropped off the action, as it should be, since the angular dynamics does not depend on  $m_0$.
In this formulation, the stochastic average have been computed exactly. To do this, we needed to introduce  four variables, 
two commuting or ``bosonic'' $(\theta,\varphi)$ and two Grassmann or ``fermionic'' ($\bar\xi,\xi$) variables. Interestingly, 
we have recently shown~\cite{arenas2012,Arenas2012-2} that this form of the action has a hidden supersymmetry that codifies 
the equilibrium properties in the form of  Ward-Takahashi identities.

We can simplify this formulation by integrating out the Grassmann variables, using the proper regularization prescription 
$\langle\bar\xi(t)\xi(t)\rangle=1/2$~\cite{arenas2010, Arenas2012-2}. The response variable $\varphi(t)$ can be also integrated 
exactly, obtaining finally,
\begin{equation}
Z(J)=\int {\cal D}\theta\;  e^{-S[\theta] + \, \int dt\,
 (J_x \cos\theta+J_y \sin\theta)} \; ,
 \label{eq.Ztheta}
\end{equation}
 with the effective action 
\begin{equation}
S[\theta] = \int dt \; \left\{\frac{1}{2g^2}\left(\frac{d\theta}{dt}-f + \frac{1}{2} g g'\right)^2 + \frac{1}{2}f'\right\}.
\label{eq.Stheta}
\end{equation}
Eqs.~(\ref{eq.Ztheta}) and~(\ref{eq.Stheta}) are the Onsager-Machlup~\cite{Onsager1953,Zinn-Justin} representation of the 
Langevin Eq.~(\ref{eq.1noise}) with the diffusion function given by Eq.~(\ref{eq.g}). 

Thus, the main result of this section is that the generating functional representation of correlation functions of the stochastic 
process driven by Eq.~(\ref{eq.sLLG}) with a vectorial noise  $\vec b$, is completely equivalent to the representation of 
correlation functions for the process driven by Eq.~(\ref{eq.1noise}), with just one scalar noise with the diffusion function 
$g(\theta)$ given  by Eq.~(\ref{eq.g}). This implies, for instance, that 
\begin{equation}
\langle m_x(t_1)\ldots m_x(t_n)\rangle_{\vec b}=m_0^n\langle \cos\theta(t_1)\ldots\cos\theta(t_n)\rangle_\eta \ ,
\end{equation}
where the first average is computed with Eq.~(\ref{eq.sLLG}) and the second one with Eq.~(\ref{eq.1noise}).

%%%%%%%%%%%%%%%%%%%%%%%%%%%%%%%
\section{Conclusions}
\label{sec.Discussion}
%%%%%%%%%%%%%%%%%%%%%%%%%%%%%%%
We have studied the dynamics of a two-dimensional local magnetization embedded in a  three-dimensional environment. We have worked in 
the ``generalized Stratonovich'' or $\alpha$-prescription in order to properly define the stochastic Landau-Lifshitz-Gilbert equation (sLLG).  
We have shown that, due to the stochastic calculus associated with each prescription (generalized chain rule of Eq.~(\ref{eq.ChainRule})), 
only the Stratonovich interpretation keeps the magnetization modulus constant  in the original formulation of the sLLG equation. Thus, 
we have presented a covariant generalization of the sLLG equation by defining an  ``$\alpha-$ covariant time  derivative''. 
This set of equations, together with the $\alpha$-calculus rules, makes the description of the stochastic process unique. 
Interestingly, the structure of the covariant derivative allows us to recognize  changes in a discretization scheme as  
``gauge transformations''. These transformations change the form of the stochastic differential equation as well as  the calculus 
rules, however, without modifying any stochastic average. 
By projecting the covariant sLLG equation to two dimensions, we showed that the dynamics is exactly described by a single Langevin 
equation for the phase variable with just one scalar multiplicative noise term. 
The character of this noise depends on the correlations between components of the original magnetic fluctuations. For uncorrelated noise, 
the effective process is a Markov additive one. However, for anisotropic and/or correlated noise, the process is truly multiplicative.  
This fact has direct consequences 
on the equilibrium probability distribution of the process. While in the case of uncorrelated noise, the distribution is of the Boltzmann 
type, in the case of anisotropic and/or correlated noise, the asymptotic distribution deviates from that behaviour.

Finally, we have built up a path integral formulation that allows us to compute, in principle, any  $n-$point correlation function.  
We have shown that the equivalence of the vectorial formulations with three correlated noises, Eq.~(\ref{eq.GsLLG}), 
and the polar formulation, with just one multiplicative noise 
(Eq.~(\ref{eq.G1noise})) is exact, \emph{i.e.}, it is valid for any $n-$point correlation function and for any value of the 
discretization scheme $0\le\alpha\le 1$.

\acknowledgments
We are in debt with Daniel A. Stariolo, for suggesting us to face the study of the sLLG equation.
The Brazilian agencies {\em Conselho Nacional de Desenvolvimento Cient\'\i fico e Tecnol\'ogico} (CNPq), {\em Funda\c c\~ao  Carlos Chagas Filho de Amparo \`a Pesquisa do Estado do Rio de Janeiro} (FAPERJ), and {\em Coordena\c c\~ao de Aperfei\c coamento de Pessoal de N\'\i vel Superior} (CAPES) are acknowledged for partial financial support.  D.G.B also acknowledges partial financial support by the  Associate Program of the Abdus Salam International Centre for Theoretical Physics, ICTP, Trieste, Italy.

%

%\bibliography{stochastic110118,ultrathin,nematic-melting,nematics-27-11-16}
\end{document}